# Dynamics of Inner Galactic Disks: The Striking Case of M100

Isaac Shlosman

Department of Physics and Astronomy, University of Kentucky
Lexington, KY 40506-0055, U.S.A.

**Abstract.** We investigate gas dynamics in the presence of a *double* inner Lindblad resonance within a barred disk galaxy. Using an example of a prominent spiral, M100, we reproduce the basic central morphology, including four dominant regions of star formation corresponding to the compression maxima in the gas. These active star forming sites delineate an inner boundary (so-called nuclear ring) of a rather broad oval detected in the near-infrared. We find that inclusion of self-gravitational effects in the gas is necessary in order to understand its behavior in the vicinity of the resonances and its subsequent evolution. The self-gravity of the gas is also crucial to estimate the effect of a massive nuclear ring on periodic orbits in the stellar bar.

# 1 Nuclear Starburst Galaxies

The role of active galaxies within the framework of galactic evolution is far from clear. In particular, it is still unknown if central activity encompasses a small percentage of galaxies or if it is a normal evolutionary stage. It also remains undetermined whether two notable types of such activity, nuclear starbursts and Seyfert nuclei, are 'genetically' linked.

A subgroup of disk galaxies shows intense star forming activity within the central few hundred parsecs, in the so-called nuclear rings. In the visual and ultraviolet wavelength ranges and at a high spatial resolution, these rings frequently appear to be patchy and incomplete, and/or consist of a pair of tightly wound spiral arms (Buta and Crocker 1993). In the near-infrared (NIR), they are regular and weakly elliptical (Knapen et al. 1995a; Shaw et al. 1995). Molecular gas distribution based on CO emission reveals a complex morphology at and interior to the rings (review by Kenney 1996). The origin of these rings is related to the inner Lindblad resonance (ILR), i.e. the resonance between the planar stellar orbits and the perturbing force of a stellar bar or of an oval distortion. Although self-gravity in the gas was ignored in earlier numerical simulations of disk galaxies, these simulations clearly showed that gas is focused into nearly circular orbits interior to the *outer* ILR (OILR; e.g. Schwarz 1984; Combes and Gerin 1985). As the gas accumulates in this region, no further evolution occurs and it was suggested that the gas is converted into stars with a high efficiency (Elmegreen 1994). More sophisticated 2D modeling involving self-gravity in the gas supported this picture (Shaw et al. 1993), indicating at the same time that increasing surface density in the gas may lead to its fragmentation and destruction of the gaseous ring (Wada and Habe 1992). Effects of fragmentation are lessened if energy deposition by massive stars is taken into account by inducing turbulence in the gas (Heller and Shlosman 1994).

Recent observations of molecular gas distribution in the centers of a number of barred galaxies provide some indirect evidence (based on stellar rotation curves) that a double ILR may in fact reside there (Kenney et al. 1992; Knapen et al. 1995a). Although gas flows in barred galaxies have been modeled for about two decades now, gas behavior in the vicinity of a double ILR is not well understood. Due to the low resolution of numerical schemes and to a common belief that the *inner* ILR (IILR) is probably located too close to the center to be observationally resolved, most efforts have been devoted to the study of gas dynamics between the OILR and outer Lindblad resonance (OLR; e.g. Athanassoula 1992). For these reasons, numerical schemes usually fail to catch the nuclear ring phenomenon.

Here we investigate a self-consistent gas evolution in the central resonance region of a moderately barred galaxy. Effects of star formation are incorporated at some level. For convenience, we choose the M100-like total mass distribution which was claimed to possess a double ILR (Arsenault et al. 1988; Knapen et al. 1995a). Furthermore, we highlight the effect that massive nuclear rings have on the dominant stellar orbits in the galactic disk, i.e. the back-reaction of the stellar component to the self-gravitating gas accumulating between the ILRs. A full account can be found in Knapen et al. (1995b) and Heller and Shlosman (1996).

## 2  Twisting of NIR Isophotes in M100

M100, the brightest (barred) spiral galaxy in Virgo, displays all the virtues of a nuclear starburst and is inclined at $30° \pm 3°$. Surprisingly, the UV/optical starburst ring formed by a tight pair of spirals, lies around $10''$ from the center ($1''$ corresponds to 83 pc at a distance of 17.1 Mpc), whereas its NIR ($2.2\mu$m) counterpart has a substantially larger radius of $\sim 18''$ and a considerable width of $\sim 16''$ (Knapen et al. 1995a). It is weakly oval (minimal ellipticity $\sim 0.13$) and its semimajor axis leads the stellar bar by about $60° - 70°$. Most unusual is the observed gradual twist of the NIR isophotes, from the position angle (P.A.) of the ring towards the P.A. of the stellar bar, both *exterior* and *interior* to the ring (see Fig. 1). In other words, the large-scale bar is dissected by an oval ringlike zone oblique to the bar. The existence of this inner bar-like feature was confirmed independently by high-resolution CO observations of the velocity field (Rand et al. 1996). Molecular gas within the central $10''$ participates in high velocity $\sim 100\,\mathrm{km\,s^{-1}}$ non-circular motions along the P.A. of the stellar bar.

**Fig.1.** The K and I isophotes in M100: position angle (*top*) and ellipticity (*bottom*) (From Knapen et al. 1995b).

Stellar orbits are expected to change their orientation abruptly by $90°$ at each resonance in the disk, which can be understood within the framework of forced oscillations. These orbits are oriented along the stellar bar between the OILR and the corotation radius ($x_1$ family), they are perpendicular to the bar between the IILR and OILR ($x_2$ family), and are again aligned with the bar between the center and the IILR. NIR isophotes in a quiescent galaxy are believed to follow the overall mass distribution, i.e. they arise from an old stellar population represented by K and M giants, (e.g. Frogel 1985). However, *gradual* skewing of isophotes in M100 is indicative of gas behavior across the resonances rather than the behavior of stars. There are two possible ways to explain this twisting of NIR isophotes between $10''$ to $30''$ in M100. First, there can be a sufficient contribution from massive young stars, K and M supergiants (Knapen et al. 1995a). These stars are expected to be found in the star bursting regions, and, being dynamically young (age less than $10^7$ yrs), they should follow gas rather than stellar orbits. In such a case, the NIR light will *not* follow the mass distribution. Knapen et al. (1995b) found that this is the most plausible explanation for two NIR 'hot spots' in M100, and similar conclusions have been reached for active star forming regions in NGC 1309 (Rhoads 1996). Alternatively, gas gravity can drag some of the old population stars in the stellar bar towards the $x_2$ orbits in the ring (Shaw et al. 1993). As we show in Section 4, the growing accumulation of gas between the ILRs is capable of affecting the main periodic orbits in the bar (especially the $x_1$ orbits outside the OILR) in such a way as to support a gradual twist of *stellar* orbits, from being aligned with the bar to becoming almost perpendicular to it.

## 3   Simulations of Stellar and Gas Dynamics in M100

To further understand the circumnuclear morphology in M100 and to confirm that it is compatible with the presense of a double ILR there, we have tailored our numerical simulations to a M100-*like* mass distribution. Stars and gas, embedded in halo and bulge potentials, were evolved by means of a 3D hybrid SPH/$N$-body code (Heller and Shlosman 1994). Models without and with star formation have been constructed, Q1 and Q2 respectively. In the Q2 model, the gas is considered to undergo 'star formation' if it is Jeans unstable, if it participates in a locally converging flow, and if its density exceeds 20 $M_\odot$ pc$^{-3}$. We assume that only massive OB stars form, which deposit $10^{51}$ ergs per $10^6$ yrs per star in the gas by means of their line-driven winds, as well as deposit $10^{51}$ ergs per $10^4$ yrs as supernovae, leaving no remnants. This energy is assumed to be instantly radiated away by the gas, with only 5% being retained and converted into turbulent motion. One unit of time, $[\tau]$, corresponds to $3.75 \times 10^7$ yrs.

Initially, an axisymmetric model of a galactic disk was constructed which was dynamically unstable and developed a stellar bar in the process of evolution. The resulting bar strength was moderate, $q \sim 0.3$, where $q$ is defined as a maximum ratio of the sum of $m = 2, 4, 6$, and 8 Fourier components of the nonaxisymmetric gravitational force to the $m = 0$ axisymmetric component. Based on the (axisymmetric) rotation curve of the model after $\sim 3$ rotation periods, a double ILR could be found in the central region, IILR at $\sim 240$ pc and OILR at $\sim 1.4$ kpc. However, the epicyclic approximation is invalid because of the relatively strong bar. Instead, we have analyzed the dominant families of periodic orbits in the model potential. It is customary to extend the ILRs into the nonlinear regime based on the radial extent of the $x_2$ family of orbits. The corresponding limiting semimajor orbital axes in Q1 and Q2 have been found at $\sim 500$ pc and $\sim 1.3$ kpc, and adopted as the nonlinear IILR and OILR, respectively. Hence, the resonance region between the ILRs was reduced substantially compared to the linear regime.

After an initial transient in the Q1 model, the gas formed a pair of trailing shocks along the leading edge of the stellar bar, in agreement with 2D numerical simulations (e.g. Athanassoula 1992). Deeper in the potential well, gas dynamics in the vicinity of ILRs was dominated by a pair of tightly wound trailing shocks and a pair of leading shocks (see Fig. 2, at $\tau = 15$). Such a shock system in the vicinity of a double ILR is predicted on the base of a simple epicyclic approximation which is not adequate at this stage. In particular, we note that two systems of shocks, trailing and leading, interact non-linearly through a pronounced cuspy feature (caustic; Fig. 2, $\tau \gtrsim 17$). To the extent that this shock system delineates spiral arms, we observe a transient pseudo-ring made out of a pair of tightly wound spirals between the ILRs. In fact, two separate gas circulations (oval 'rings') form between the ILRs due to the action of the shocks. The outer gaseous ring evolves as to (almost) align itself with the minor axis of the stellar bar, while the inner gaseous ring (almost) aligns itself with the major axis of the bar.

This evolution comes about as a response to the gravitational torques by the bar and in order to minimize them. However, the effects of self-gravity in the gas are crucial in order to understand its dynamics. The self-gravity acts in a manner similar to surface tension: the outer gaseous ring settles down on lower energy $x_2$ orbits, away from the OILR. At the same time, the inner ring shrinks across the IILR and settles on the $x_1$ orbits. This behavior is depicted in Figure 2, at around $\tau \sim 17-24$. It is imperative to mention that the gas is *not* locked between the ILRs, but merely experiences a temporary slowdown there. The outer ring is constantly perturbed by a number of density inhomogeneities and 'rains' onto the inner ring which contains orbits deeper in the potential well. (In the presence of star formation, as we discuss below, these perturbations are caused by the turbulent motions in the gas excited by stellar winds and supernovae.) As the inner ring becomes more massive, it is subject to gravitational fragmentation. Neglecting the effects of star formation at this stage is not justified.

Technically speaking, star formation in the Q2 model has introduced 'turbulent' motion in the gas and induced mixing between material with a different angular momentum and energy. As a result, gaseous circulations between the ILRs widen notably and merge, and it is more appropriate to speak of a gaseous disk which extends inwards, from the outermost $x_2$ orbits, much of the way to the center. This disk is an oval, and its semimajor axis is positioned almost at right angles to the bar.

As expected, the shock system outlines the regions of intense star formation in the Q2 model (see Fig. 3). However, there is no one-to-one correspondence between the shocked and gravitationally unstable gas, as the shock and star formation maps show. At the same time, four major sites of star formation in the resonance region persist during most of the simulation time — all corresponding to the maxima of dissipation in the gas. Two elongated star forming regions are found downstream from the place where the inflow along the outer shocks crosses the bar's minor axis and encounters the gas circulation on the $x_2$ orbits. Azimuthal smearing of star forming regions is a direct consequence of the time scale for Jeans instability becoming an appreciable fraction of the orbital time scale so close to the rotation axis. Another pair of star forming regions is located around the IILR ($\tau \sim 10-25$, in Fig. 3), slightly ahead of the bar's major axis and where the cuspy feature is seen in the Q1 model. This prevailing morphology dominated by four star forming regions (at and just outside the IILR) appears to be robust during most of the simulation time. It is compatible with the loci of star formation in the $U$, $V$, and H$\alpha$ images of the inner 2 kpc in M100 (Knapen et al. 1995a). No star formation correlates with the position of the OILR.

**Fig.2.** Logarithmic gray-scale map of shock dissipation inside the OILR in M100 (Q1 model, without star formation). Time is given in the upper left corners, $[\tau = 1] = 3.75 \times 10^7$ yrs. Each frame is 2.6 kpc across. The gas flows counterclockwise, the stellar bar is horizontal. (From Knapen et al. 1995b).

The star formation rate in the Q2 model reaches its peak around $\tau \sim 25-28$, exhibiting burst behavior with a typical time scale of $\sim 10^7$ yrs. Around $\tau \sim 28$, the mass inflow rate across the IILR peaks strongly, indicating a catastrophic loss of angular momentum by the gas ($\tau = 28.4$, Fig. 3). The subsequent evolution of this dynamical runaway in the self-gravitating gas was discussed by Heller and Shlosman (1994) in the absence of star formation. We view this process as the initial phase of decoupling of the gaseous bar from the large-scale stellar bar, as envisioned in the 'bars within bars' scenario (Shlosman, Frank and Begelman 1989). It is not clear how much of the underlying old population participates in this process (and if it does at all). The amount of gas at the onset of instability, $\sim$few$\times 10^9$ M$_\odot$, is $\sim 10\% - 20\%$ of the mass *interior* to the runaway region, which is more than an order of magnitude less than quoted by Ho, Filippenko and Sargent (1996). This amount of gas is available even in the early-type disks.

In the more realistic case, advanced here in the Q2 model, the outcome of evolution depends in a sensitive way on the efficiency of star formation, the state of the interstellar medium, and its ability to retain energy deposited by massive stars. These questions are potentially relevant for our understanding of fueling the nonthermal activity in Seyfert nuclei and must be addressed in future work. We conclude, that a characteristic time for the gas to 'filter' through the resonance region is $\lesssim 10^9$ yrs. This can be taken as a rough estimate of a nuclear ring's lifetime.

## 4    Massive Nuclear Rings: Affecting Periodic Orbits

The above numerical simulation of gas dynamics in the presence of a double ILR demonstrates explicitly that the radial gas inflow is not stopped between the resonances but only slows down there — a kind of 'self-mulching lawn mower' effect. This results in gas accumulation in the form of a massive elliptical 'ring'. Molecular mass of as much as a few$\times 10^9$ M$_\odot$ stored in the ring is not out of the question. Besides enhanced star formation caused by favorable conditions in the region, such a ring will have gravitational effects on the stellar component in the disk. Here, we are mainly interested in how the main periodic orbits are affected, and what consequences it may have on the gas circulation in the bar.

For simplicity we use a 3D analytical model of a galaxy consisting of a disk, bulge, halo and stellar bar. Furthermore, in an attempt to simulate the effect of a massive nuclear ring within the central kpc, we make assumptions about its shape and the orientation of its major axis with respect to the stellar bar. Results of a 3D orbit analysis for such gravitational potentials are presented elsewhere (Heller and Shlosman 1996). Here we discuss only necessary details.

**Fig.3.** *Left:* Logarithmic gray-scale map of shock dissipation inside the OILR in M100 (model with star formation). Time is given in the upper left corners, $[\tau = 1] = 3.75 \times 10^7$ yrs. Each frame is 2.6 kpc across. The gas flows counterclockwise, the stellar bar is horizontal. *Right:* Star formation map corresponding to the region shown on the left. (From Knapen et al. 1995b).

The most pronounced change in the periodic orbits, when a circular ring (or elliptical ring whose major axis coincides with that of the bar) is added, is that the extent of the ILR resonance region is increased with the ring's mass, weakening the stellar bar. In addition, the $x_1$ orbits of different energy intersect in the vicinity of the ring (as do $x_2$ orbits within the ring). This has a two-fold effect on the gas: the phase space available to $x_2$ has increased, but at the same time, orbits near the ring became intersecting and unable to hold gas, amplifying dissipation there. Thus, we expect that the gas will fall through the IILR after an initial stage of accumulation, exactly as observed in the numerical simulations.

An additional and qualitatively different effect is obtained when the ring is mildly elliptical and its major axis is oblique to the bar, leading it in the direction of galactic rotation. This configuration is the one typically observed in nuclear starburst galaxies and appears as a long-lived transient in our numerical simulations. The behavior of $x_1$ orbits can be qualitatively understood in this case as a response to the perturbing forces of the stellar bar and of the oblique ring. Both forces have the same driving frequency but are phase-shifted. A straightforward application of an epicyclic approximation to the motion of a viscous 'fluid' reveals the rich variety of possible responses in the gas to this driving force. A representative case, calculated using fully nonlinear orbit analysis when the ring leads the stellar bar by 60°, is shown in Figure 4. The change in the position angle of the $x_1$ with distance to the ring is rather dramatic, starting at large radii with a slowly growing phase shift which reaches a maximum of 34° in the leading direction, followed by a rapid decline to 0° just interior to the ring and then continuing to –11° in the trailing direction deep into the bulge.

**Fig. 4.** Twisting of $x_1$ periodic orbits supporting the stellar bar in the presence of a massive $10^9$ $M_\odot$ oblique elliptical ring ($e = 0.4$). Ring's potential is softened with $\epsilon = 100$ pc. The stellar bar is horizontal. Rotation is counterclockwise and corotation is at 5 kpc. The ring's semimajor axis ($[r = 0.04] = 400$ pc) is offset to the bar by 60° in the leading direction. The frame is 2 kpc on a side. (From Heller and Shlosman 1996).

## 5   Implications: Dynamics of the Circumnuclear Region

Our dynamical study of gas flow across a double ILR resonance in the circumnuclear region of a barred galaxy provides some insight into relevant processes which accompany this flow. First, self-gravitating effects in the gas accumulating between the resonances are crucial in understanding its subsequent evolution (besides the star formation). In particular, self-gravity acts as a 'surface tension' and the gas moves deeper into the potential well, away from the OILR and across the IILR. Star formation is clearly peaked around the IILR, partly because the gravitational torque's sign changes across this resonance causing additional compression in the gas and creating conditions favorable to Jeans instability there.

Second, we find a complicated but basically a ring-like morphology for the distribution of star forming regions in the nuclear starbursts. Dominant regions of star formation lie downstream from two main compression sites of the gas ('twin peaks'), when it crosses the minor axis of the stellar bar. This is somewhat outside the IILR. Additional pair of star forming regions are found at the IILR, on the major axis of the bar and slightly leading it, and correspond to the twin 'hot spots' found in M100.

Third, there is no prominent star formation associated with the OILR. This explains why the UV/optical and H$\alpha$ observations identify the star forming ring in M100 *inside* its NIR ring. Unless the star formation, already at this stage, is very efficient in consuming the gas at the IILR, the gas 'breaks through' and falls towards the center, an event that is accompanied by a prominent burst of star formation. This dynamical runaway of self-gravitating gas *inside* the IILR depends on details of star formation and physics of the interstellar medium, and it is outside the scope of this study. We only comment that the characteristic time scale of this runaway is $\sim$few$\times 10^7$ yrs, much shorter than the time it takes for the gas to 'filter' through the resonance region, $\lesssim 10^9$ yrs.

Fourth, we find that massive nuclear rings are capable of perturbing the gravitational potential in the circumnuclear regions, thus affecting the main periodic orbits there. The phase space allowed to the orbits aligned with the minor axis of the bar ($x_2$ family) is substantially increased and the orbits aligned with the bar ($x_1$ family) are significantly distorted. Orbits with different values of the Jacobi integral are found to intersect, meaning that gas cannot be retained there and will move inwards across the IILR. In the most interesting case of an elliptical ring oblique to the bar, $x_1$ orbits are gradually twisted, in a way similar to the skewing of NIR isophotes observed in M100. So these orbits can trap both gas (accompanied by K and M supergiants) and old population stars. It may be possible, in a such a case, to estimate the 'degree of asymmetry' in the nuclear ring (i.e. azimuthal mass distribution) based on the observed change in the ellipticity and position angle of NIR isophotes with radius.

To summarize, the gas seems to be the prime dynamic agent in the circumnuclear regions of at least some disk galaxies, although its mass is only a fraction of the mass in the stellar component there. Further observations of molecular gas kinematics in active and normal galaxies will provide clues to understanding their central activity.

ACKNOWLEDGEMENTS. It is a pleasure to thank organizers of this stimulating meeting for financial assistance. I am indebted to John Beckman, Roelof de Jong, Clayton Heller, Johan Knapen and Reynier Peletier for collaboration on some of the research described above.

## References

Arsenault, R., Boulesteix, J., Georgelin, Y., Roy, J.-R. (1988): A&A **200**, 29
Athanassoula, E. (1992): MNRAS **259**, 345
Buta, R., Crocker, D.A. (1993): AJ **105**, 1344
Combes, F., Gerin, M. (1985): A&A **150**, 327
Elmegreen, B.G. (1994): ApJL **425**, L73
Frogel, J.A. (1985): ApJ **298**, 528
Heller, C.H., Shlosman, I. (1994): ApJ **424**, 84
Heller, C.H., Shlosman, I. (1996): ApJ, submitted
Ho, L.C., Filippenko, A.V., Sargent, W.L.W. (1996): IAU Colloq. 157 on *Barred Galaxies* (Kluwer, Dordrecht, Boston, London), eds. R. Buta et al., in press
Kenney, J.D.P. (1996): *The Interstellar Medium in Galaxies* (Kluwer, Dordrecht, Boston, London), ed. J.M. van der Hulst, in press
Kenney, J.D.P., Wilson, C.D., Scoville, N.Z., Devereux, N.A., Young, J.S. (1992): ApJL **395**, L79
Knapen, J.H., Beckman, J.E., Heller, C.H., Shlosman, de Jong, R.S. (1995b): ApJ **454**, 623
Knapen, J.H., Beckman, J.E., Shlosman, I., Peletier, R.F., Heller, C.H., de Jong, R.S. (1995a): ApJL **443**, L43
Rand et al. (1996): in preparation
Rhoads, J.E. (1996): ApJL, submitted
Schwarz, M.P. (1984): MNRAS **209**, 93
Shaw, M., Axon, D., Probst, R., Gatley, I. (1995): MNRAS **274**, 369
Shaw, M., Combes, F., Axon, D.J., Wright, G.S. (1993): A&A **273**, 31
Shlosman, I., Frank, J., Begelman, M.C. (1989): Nature **338**, 45
Wada, K., Habe, A. (1992): MNRAS **259**, 82